\documentclass[11pt]{article}

\usepackage[T1]{fontenc}
\usepackage[utf8]{inputenc}
\usepackage{lmodern}

\usepackage{amsmath,amssymb,bm}
\usepackage{graphicx}
\usepackage{booktabs}
\usepackage{array}
\usepackage{microtype}
\usepackage{geometry}
\usepackage{hyperref}
\usepackage{threeparttable}
\usepackage[backend=biber,style=numeric,sorting=none]{biblatex}
\usepackage{caption}
\usepackage{siunitx}
\usepackage{authblk}

\hypersetup{colorlinks=true, citecolor=blue, linkcolor=blue, urlcolor=blue}
\title{Equation of State of the Quark-Gluon Plasma in an External Magnetic Field to Fourth Order}
\author[1]{L. Levkova\footnote{Corresponding author, email: lal43@caa.columbia.edu}}
\author[2]{C. DeTar}
\affil[1]{Independent Researcher, Bethesda, Maryland 20815, USA}
\affil[2]{Department of Physics and Astronomy, University of Utah, Salt Lake City, Utah 84112, USA}
\date{}

\begin{document}
\maketitle

\begin{abstract}
We extend a lattice-QCD determination of the response of the quark-gluon plasma to an external magnetic field using a Taylor expansion of the pressure at zero field. A zero-net-flux ``half-and-half'' magnetic-field configuration permits continuous differentiation with respect to the field amplitude on gauge ensembles generated at $B=0$. The electromagnetic vector potential is chosen to minimize stochastic noise in the Taylor operators. Previous results
    through $O(B^2)$ are extended to $O(B^4)$ using $2+1$-flavor HISQ/tree ensembles with $m_l/m_s=0.05$ and $N_t=8$. The renormalized quadratic coefficient confirms paramagnetic behavior above the QCD crossover. The fourth-order coefficient 
    $C_4$ is small at high temperature and is enhanced near the crossover. At $T=167$ MeV the $C_4$ result at $48^3\times 8$ lies about two standard deviations above the corresponding $32^3\times8$ one, indicating likely finite-volume effects. For $eB = 0.2$ GeV$^2$, characteristic of magnetic fields relevant to heavy-ion phenomenology, the fourth-order contribution is small compared with the leading magnetic correction at moderate and high temperatures, but near the crossover the
    large-volume result is about half of the central value of the quadratic term, after the vacuum-pressure subtraction. We also place the method in the context of other lattice determinations of magnetic susceptibility and clarify the relevance of finite-volume effects for the thermal subtraction. This work was completed in 2014 and preliminary results were reported at Lattice 2014.
\end{abstract}

\section{Introduction}

Large magnetic fields occur in several environments in which strongly interacting matter is important. Fields of order $10^{15}$~T may be generated transiently in noncentral relativistic heavy-ion collisions, while still larger primordial fields have been considered in early-Universe scenarios. Magnetic fields can modify the thermodynamics of QCD matter and therefore may influence the evolution of the quark-gluon plasma and other strongly interacting systems \cite{LevkovaDeTar2014,Endrodi2013HRG,Bali2014EOS}.

Lattice QCD provides a first-principles framework for determining this response. However, a complication arises because a constant magnetic field on a periodic spatial torus carries quantized magnetic flux \cite{AlHashimiWiese2009}. Consequently, the field strength cannot normally be varied continuously, making derivatives of the partition function with respect to $B$ inconvenient. In Ref.~\cite{LevkovaDeTar2014}, we introduced a zero-net-flux magnetic-field profile that is positive on one half
of the lattice and negative on the other. The amplitude $B$ of the magnetic field is continuous, so derivatives with respect to it can be evaluated directly at $B=0$. This enables a Taylor expansion analogous to the standard chemical-potential expansion and avoids generating separate dynamical ensembles for every value of $B$.

The original calculation determined the magnetic contribution to the pressure through $O(B^2)$ and found that the thermal QCD medium is paramagnetic above the crossover \cite{LevkovaDeTar2014}. The extension presented here carries the same framework to $O(B^4)$. The purpose is twofold: to determine the leading nonlinear magnetic response and to test how far the quadratic approximation remains adequate for phenomenologically relevant fields.
In the following we use the same notations and conventions as in Ref.~\cite{LevkovaDeTar2014}.

\section{Taylor expansion and magnetic-field construction}

The pressure is expanded in powers of the dimensionless magnetic field as
\begin{equation}
\frac{p(T,B)}{T^4}
=\frac{\ln Z(B)}{T^3V}
=\sum_{n=0}^{\infty} C_n(T)
\left(\frac{|e|B}{T^2}\right)^n,
\end{equation}
with
\begin{equation}
C_n(T)=\frac{L_t^3}{L_s^3}\frac{1}{n!}
\left.
\frac{\partial^n\ln Z}
{\partial(|e|B/T^2)^n}
\right|_{B=0}.
\end{equation}
Because the pressure is CP-even, only even coefficients contribute. The central quantities in this work are $C_2$ and $C_4$.

\subsection{Half-and-half magnetic field}

For a homogeneous magnetic field in the $z$ direction on a periodic lattice, flux quantization requires
\begin{equation}
|q|B=\frac{2\pi b}{L_xL_ya^2},\qquad b\in\mathbb{Z},
\end{equation}
where the smallest quark charge is $|q|=|e|/3$. To avoid differentiating a quantized field, we use instead a zero-net-flux profile
\begin{equation}
B_z(x)=
\begin{cases}
+B, & x\le L_x/2,\\
-B, & x > L_x/2.
\end{cases}
\end{equation}
The flux entering one half of the $x$--$y$ surface exits through the other half. The field amplitude $B$ is therefore continuous and Taylor differentiation at $B=0$ is straightforward. The sign reversals introduce surface effects, so the finite-volume behavior must be checked; for the second-order thermal quantity such effects are not discernible at the level of the available statistics as reported below. 
In contrast, the corresponding fourth-order comparison at $T=167$ MeV shows a likely volume dependence and therefore warrants a more cautious interpretation.

\subsection{Minimal-noise vector potential}

The same half-and-half magnetic field can be represented by different U(1) link phases. This freedom matters numerically because the individual stochastic contributions to the determinant derivatives depend on the magnitude and spatial distribution of those phases. 
The calculation in  Ref.~\cite{LevkovaDeTar2014} uses the symmetric choice
\begin{align}
u_y(B,q,x)&=\exp\!\left[i a^2qB(x-L_x/4)\right], && x\le L_x/2,\\
u_y(B,q,x)&=\exp\!\left[i a^2qB(3L_x/4-x)\right], && x> L_x/2,\\
u_x&=u_z=u_t=1.
\end{align}
This representative minimizes the typical phase magnitude while retaining the desired magnetic profile. Direct tests showed that the statistical noise of the Taylor observables depends strongly on this choice and that the centered configuration reduces the standard deviation by a sizable factor relative to other allowed choices. Thus the link construction functions not only as a way to circumvent flux quantization but also as an explicit variance-reduction strategy \cite{LevkovaDeTar2014}.

\subsection{Explicit form for the Taylor expansion coefficients for the pressure}
For $2+1$ rooted staggered flavors $u$, $d$ and $s$, we write the partition function as: 
\begin{equation}
Z(B)=\int dU\,e^{-S_g}e^Ue^De^S,
\end{equation}
where
\begin{equation}
U=\frac14\ln\det M_u(B,q_u),\quad
D=\frac14\ln\det M_d(B,q_d),\quad
S=\frac14\ln\det M_s(B,q_s),
\end{equation}
with $M_{u,d,s}$ being the HISQ fermion matrices for the corresponding quark flavors and $S_g$ is the gluon action.
Defining $A_{nml}$ as the appropriately charge-normalized expectation value of derivatives of $e^U$, $e^D$, and $e^S$:
\begin{equation}
    A_{nml}=
    \frac{1}{q_u^nq_d^mq_s^l}
    \left\langle
    e^{-U-D-S}
    \frac{\partial^n e^U}{\partial(a^2B)^n}
    \frac{\partial^m e^D}{\partial(a^2B)^m}
    \frac{\partial^l e^S}{\partial(a^2B)^l}
    \right\rangle, 
\end{equation}
then the second- and fourth-order coefficients are:
\begin{align}
    C_2=&\frac{1}{2L_tL_s^3}
    \left[
    (q_u^2+q_d^2)A_{200}+q_s^2A_{002}
    +2q_uq_dA_{110}+2(q_u+q_d)q_sA_{101}
    \right],\\
C_4=&\frac{1}{4!L_t^5L_s^3}
        \Big[
            (q_u^4+q_d^4)A_{400}
            +q_s^4A_{004}
            +12q_uq_dq_s
            (q_sA_{112}+(q_u+q_d)A_{121})
            +6(q_u^2q_d^2A_{220}\nonumber\\
            &+(q_u^2+q_d^2)q_s^2A_{022})
            +4((q_u^3q_d+q_d^3q_u)A_{310}
            +(q_u^3+q_d^3)q_sA_{301}
            +(q_u+q_d)q_s^3A_{103})\nonumber\\
            &-3(2L_tL_s^3C_2)^2
            \Big].
\end{align}
The required  $A_{nml}$ are computed using Gaussian stochastic estimators. Fourth order contributions are substantially more expensive than the second order ones because they require higher determinant derivatives and a larger set of noisy contractions.

\section{Renormalization and thermal pressure}

The magnetic field modifies both the thermal medium and the QCD vacuum. To isolate the thermal contribution to the pressure, define
\begin{equation}
    \Delta p(B,T)=p(B,T)-p(0,T)-p(B,0)+p(0,0) = C_2^r (eB)^2 + C_4^r (eB)^4/T^4 + \dots .
\end{equation}
For the above Taylor coefficients we have:
\begin{equation}
C_n^r(T)=C_n(T)-C_n(0).
\end{equation}
At $O(B^2)$, the zero-temperature subtraction removes the ultraviolet divergence associated with electric-charge renormalization, so $C_2^r$ is entirely thermal in this prescription \cite{LevkovaDeTar2014,Endrodi2013HRG}. If the conventional susceptibility is defined by
\begin{equation}
\Delta p=\frac12\chi(eB)^2+\cdots,
\end{equation}
then
\begin{equation}
\chi=2C_2^r.
\end{equation}

For $C_4$, no analogous charge renormalization divergence is present, however, without the zero-temperature subtraction, it contains both thermal and vacuum contributions. The numerical results in Table~\ref{tab:ensembles} show $C_4$ without a zero-temperature subtraction. This distinction is important: $C_2^r$ is a renormalized thermal coefficient, whereas the tabulated $C_4$ should not be relabeled $C_4^r$ without the corresponding zero-temperature subtraction.

\section{Lattice calculation}

We use $2+1$-flavor HotQCD HISQ/tree ensembles along the line of constant physics $m_l/m_s=0.05$ \cite{Bazavov2012HotQCD}. The non-zero temperature lattices have $N_t=8$ and cover temperatures from 134 to 611~MeV. Matching zero-temperature ensembles are used for the second-order subtraction. The previous $O(B^2)$ analysis \cite{LevkovaDeTar2014} was extended to $O(B^4)$ and the non-zero temperature statistics were substantially increased. The extended calculation consumed
more than $4.4\times10^5$ GPU-hours on 2014-era Fermi/Kepler-class GPUs\footnote{The computational cost would be roughly $4\times10^4$ H100 GPU-hours, before accounting for subsequent improvements in QUDA.} with QUDA \cite{Clark2010QUDA,Babich2011QUDA}, with more than 20\% of the cost devoted to finite-volume checks. This total should not be interpreted as the cost of $C_2$ alone: the $O(B^4)$ operators are substantially more expensive and much noisier.
\begin{table}[htbp]
        \centering
        \caption{Lattice parameters and Taylor coefficients. $C_2^r$ is given in units of $10^{-3}$ and $C_4$ in units of $10^{-5}$.}
        \label{tab:ensembles}

        \resizebox{\textwidth}{!}{%
             \begin{threeparttable}
            \begin{tabular}{ccccccccc}
                \toprule
                $T$ [MeV] & $\beta$ & $m_l/m_s$ & $V_T$ & $V_0$ &
                sources $T/0$ & configs $T/0$ & $10^3 C_2^r$ & $10^5 C_4$\\
                \midrule
                134 & 6.195 & 0.00440/0.0880 & $32^3\!\times8$ & $32^3\!\times32$ &
                4800/400 & 210/50 & $-0.4(4)$ & $1.7(1.0)$\\
                154 & 6.341 & 0.00370/0.0740 & $32^3\!\times8$ & $32^3\!\times32$ &
                4800/500 & 200/50 & $0.6(4)$ & $2.6(1.3)$\\
                167 & 6.423 & 0.00335/0.0670 & $32^3\!\times8$ & $32^3\!\times32$ &
                2400/200 & 420/50 & $2.4(5)$ & $2.5(1.3)$\\
                167 & 6.423 & 0.00335/0.0670 & $48^3\!\times8$ & $48^3\!\times48$ &
                4800,9600--28800\tnote{1}\,/400 & 150,185\tnote{2}\,/70 & $2.5(4)$ & $6.4(1.4)$\\
                173 & 6.460 & 0.00320/0.0640 & $32^3\!\times8$ & $32^3\!\times64$ &
                2400/200 & 100/60 & $3.4(5)$ & $2.8(1.1)$\\
                227 & 6.740 & 0.00238/0.0476 & $32^3\!\times8$ & $48^3\!\times48$ &
                2400/200 & 50/50 & $10.3(8)$ & $0.82(9)$\\
                373 & 7.280 & 0.00142/0.0284 & $32^3\!\times8$ & $48^3\!\times64$ &
                1200/40 & 50/50 & $19.3(1.3)$ & $0.64(5)$\\
                611 & 7.825 & 0.00082/0.0164 & $32^3\!\times8$ & $64^3\!\times64$ &
                1200/40 & 50/50 & $27.7(1.4)$ & $0.51(2)$\\
                \bottomrule
            \end{tabular}

            \begin{tablenotes}
            \item[1] For $C_2^r$ the number of sources per gauge configuration was 4800 at non-zero temperature, while for $C_4$ this number varied between 9600 and 28800.
            \item[2] The number of gauge configurations at non-zero temperature used for the determination of $C_2$ was 150, while for $C_4$ the corresponding number was 185.
            \end{tablenotes}
    \end{threeparttable}}
\end{table}

\section{Results}

\subsection{Second-order response and magnetic susceptibility}

\begin{figure}[htbp]
\centering
\includegraphics[width=0.92\textwidth]{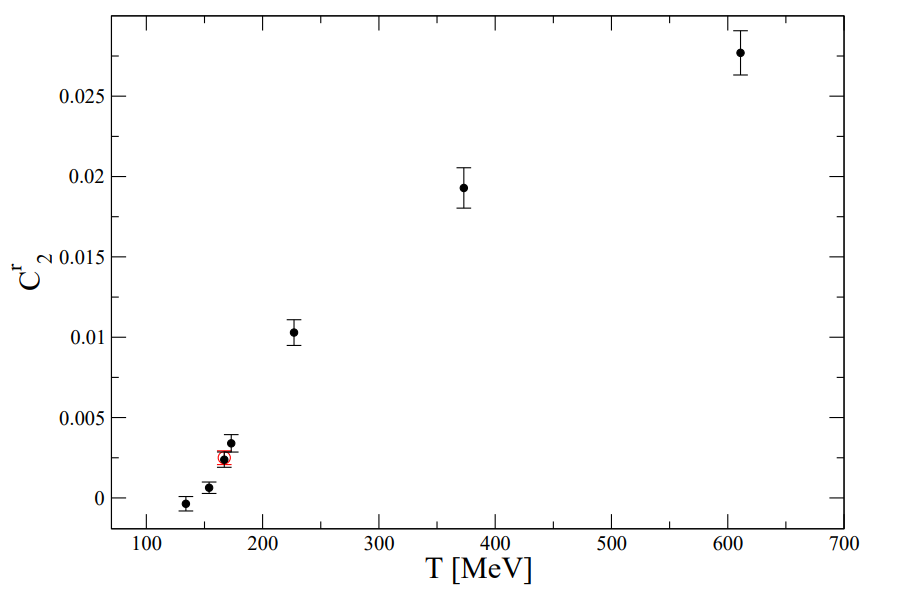}
\caption{Renormalized second-order Taylor coefficient $C_2^r$ versus temperature. The red open circle is the larger-volume result at $T=167$~MeV; the black point at the same temperature is the $32^3\times8$ result. The two volumes are statistically compatible. The change from values consistent with zero at low temperature to a clearly positive coefficient above the crossover demonstrates the onset of paramagnetic response, while the agreement of the two 167-MeV points shows no resolved finite-volume shift in $C_2^r$ at the precision of this calculation.}
\label{fig:C2r}
\end{figure}

Figure~\ref{fig:C2r} shows the renormalized second-order coefficient $C_2^r$ as a function of temperature. The two lowest-temperature values are statistically compatible with zero. Above the crossover region $C_2^r$ becomes clearly positive and rises rapidly with temperature. In the conventional normalization $\chi=2C_2^r$, this establishes a positive thermal magnetic susceptibility and therefore paramagnetic behavior of the quark-gluon plasma above the transition temperatures. 

The finite-volume test is particularly relevant to the half-and-half construction. At 167~MeV the $32^3\times8$ lattice gives $C_2^r=2.4(5)\times10^{-3}$, while the $48^3\times8$ lattice gives $2.5(4)\times10^{-3}$. The difference is far below the combined statistical uncertainty. Thus the surface effect introduced by the field reversal does not produce a detectable bias in the renormalized second-order thermal coefficient in this test. Since we expect the finite-volume effects to be strongest
in the transition region (around 167~MeV), this test may bound these effects at other temperatures.

\subsection{Fourth-order response}
\begin{figure}[htbp]
\centering
\includegraphics[width=0.92\textwidth]{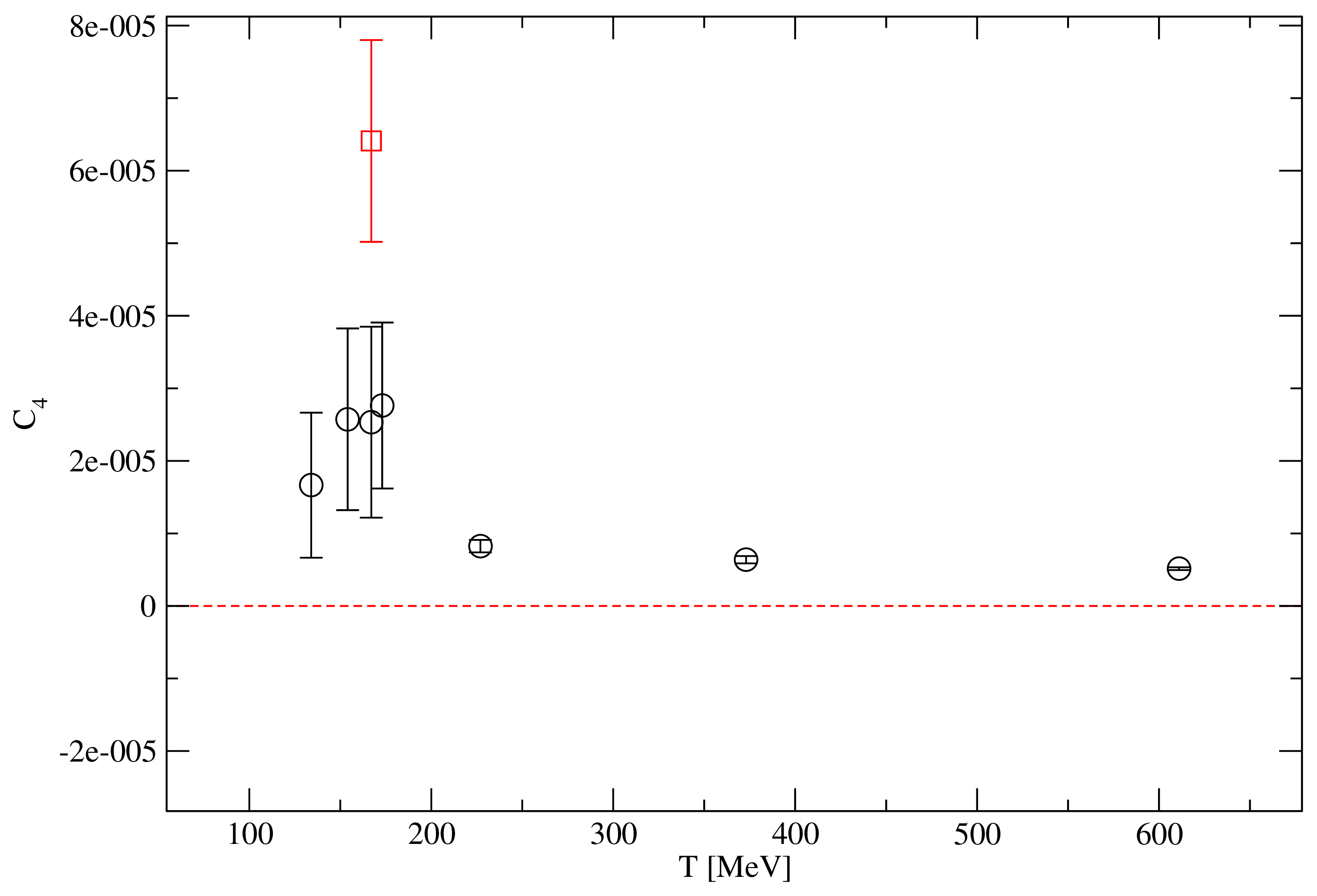}
    \caption{Fourth-order coefficient $C_4$ versus temperature. The black open circles denote the primary $32^3\times8$ non-zero temperature series, while the red open square denotes the $48^3\times8$ result at $T=167$ MeV. At this temperature the two determinations are $C_4=2.5(1.3)\times10^{-5}$ and $C_4=6.4(1.4)\times10^{-5}$, respectively. Their difference is about $2.0\sigma$, suggesting that the fourth-order coefficient may have appreciably stronger finite-volume dependence
    than $C_2^r$. At high temperature $C_4$ becomes small and is determined with good statistical precision.} 
\label{fig:C4}
\end{figure}

Figure~\ref{fig:C4} displays the fourth-order coefficient $C_4$. The central values on the $32^3\times8$ ensembles increase from $C_4=1.7(1.0)\times10^{-5}$ at $134$ MeV to values of approximately $(2$--$3)\times10^{-5}$ in the crossover region, and then decrease toward small, precisely resolved values at high temperature. The large-volume result quantifies the volume dependence at $T=167$ MeV. On the $32^3\times8$ lattice we obtain $
C_4=2.5(1.3)\times10^{-5}$,
whereas on the $48^3\times8$ lattice the result is $ C_4=6.4(1.4)\times10^{-5}$. The difference in the central values is approximately a $2.0\sigma$ effect. Thus the large-volume result provides evidence for finite-volume dependence of $C_4$ near the crossover. In contrast, the corresponding volume comparison for $C_2^r$ shows no statistically significant finite-volume effects, which we attribute to a cancellation induced by the zero-temperature subtraction. The large-volume value of $C_4$ itself is
clearly positive, differing from zero by approximately $4.6\sigma$. Nevertheless, because only two spatial volumes are available at this temperature, these data are not sufficient to determine the infinite-volume value of $C_4$ or to establish the detailed shape of its temperature dependence near the crossover.

The volume comparison emphasizes both the usefulness and the limitations of the fourth-order Taylor calculation. At high temperature $C_4$ is small and well resolved, and the quadratic approximation is therefore robust for moderate magnetic fields. Near the crossover, however, the larger $48^3\times8$ value indicates that finite-volume control becomes more important at fourth order. This behavior is consistent with the original expectation that the finite-volume effects associated
with the half-and-half construction can become more pronounced for higher Taylor coefficients. At lower temperatures the expansion parameter $eB/T^2$ also increases, so even a numerically small $C_4$ can make an increasingly important contribution to the pressure. Thus a quantitative determination of the fourth-order contribution near the crossover ultimately requires additional spatial volumes, in addition to improved statistics.

The fourth-order coefficient also has a direct interpretation as the
leading nonlinear magnetic response of the medium.  With the convention
used here, the corresponding fourth-order magnetic susceptibility is
proportional to $C_4/T^4$.  Higher derivatives of the free energy can
be particularly sensitive to changes in the relevant degrees of freedom
near a crossover, so a sufficiently precise determination of
$C_4^r(T)$ over a dense set of temperatures could provide
an additional magnetic definition of the pseudocritical temperature,
for example through the position of a maximum 
in the nonlinear response.  Such an interpretation would require both
finite-volume and continuum control.

It is also important that the $C_4$ values reported here are not
zero-temperature subtracted.  If the enhanced
finite-volume dependence observed for the half-and-half construction
is dominated by vacuum or surface contributions that are similar at
$T=0$ and finite temperature, these effects could cancel substantially
in the thermal combination $C_4^r(T,L)=C_4(T,L)-C_4(0,L)$.
Whether such a cancellation occurs at fourth order cannot be determined
from the present data, since the required zero-temperature $C_4$
measurements were not performed.  A matched $T=0$ calculation on the
two volumes at $\beta=6.423$ would therefore be especially useful: it
would distinguish a predominantly vacuum/surface finite-volume
effect from a genuine finite-volume dependence of the thermal nonlinear
magnetic response.
\subsection{Pressure at $eB=0.2\,\mathrm{GeV}^2$}

\begin{figure}[htbp]
\centering
\includegraphics[width=0.92\textwidth]{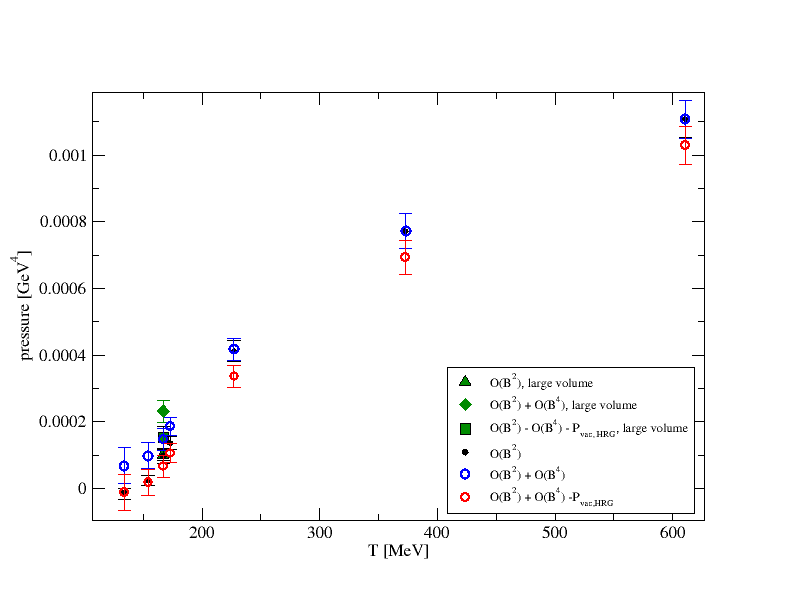}
    \caption{Magnetic contribution to the pressure at $eB=0.2~{\rm GeV}^2$. Black filled circles show the $O(B^2)$ fully thermal contribution, blue open circles show $O(B^2)+O(B^4)$ including the fourth-order vacuum contribution, and red open circles show $O(B^2)+O(B^4)$ after subtraction of the HRG estimate of the magnetic vacuum pressure. The green symbols at $T=167$ MeV denote the corresponding results obtained on the larger $48^3\times8$ lattice using both the large-volume $C_2^r$ and
    the  large-volume $C_4$: the triangle denotes $O(B^2)$, the diamond denotes $O(B^2)+O(B^4)$, and the square denotes $O(B^2)+O(B^4)-P_{\rm vac,HRG}$. The large-volume $O(B^2)$ result is statistically consistent with the $32^3\times8$ result, whereas the quantities containing the fourth-order contribution are shifted upward, reflecting the stronger volume dependence observed in $C_4$. At higher temperatures the fourth-order correction is small, while near the crossover
    the large-volume result shows that it can be appreciable. The vacuum-pressure estimate follows Ref.~\cite{Endrodi2013HRG}.}
\label{fig:pressure}
\end{figure}

Figure~\ref{fig:pressure} and Table~\ref{tab:pressure} summarize the magnetic contribution to the pressure at $eB=0.2~{\rm GeV}^2$. The $O(B^2)$ contribution is determined by the renormalized coefficient $C_2^r$. Adding the lattice $C_4$ gives the result through fourth order, while subtraction of the HRG estimate of the magnetic vacuum pressure isolates, within this prescription, the corresponding thermal contribution. The black filled circles in Fig.~\ref{fig:pressure} show the
quadratic contribution. The blue open circles include the fourth-order term before subtraction of the fourth-order vacuum contribution, and the red open circles show the result after the HRG vacuum-pressure subtraction, which removes the vacuum contribution beyond quadratic order in $B$. Table~\ref{tab:pressure} lists the numerical values corresponding to these points. The large-volume data at $T=167$ MeV provide an especially useful comparison. On the $32^3\times8$ lattice the three pressure estimates, in units of $10^{-4}~{\rm GeV}^4$, are
$\Delta p_{O(B^2)}=0.96(20)$, $\Delta p_{O(B^2)+O(B^4)}=1.47(33)$, and $ \Delta p_{O(B^2)+O(B^4)-P_{\rm vac,HRG}}=0.67(33)$. Using instead the large-volume coefficients obtained on the $48^3\times8$ lattice gives $1.00(16)$, $2.32(33)$, and $1.52(33)$, respectively. The quadratic pressure contribution is therefore essentially unchanged by the increase in volume, in agreement with the
direct comparison of $C_2^r$ in Fig.~\ref{fig:C2r}. In contrast, inclusion of the large-volume $C_4$ increases the fourth-order-inclusive pressure by approximately $0.85\times10^{-4}~{\rm GeV}^4$ relative to the corresponding $32^3\times8$ result, both before and after the common HRG vacuum-pressure subtraction. This shift is driven almost entirely by the volume dependence of $C_4$, rather than by the quadratic susceptibility. The effect is quantitatively important near the
crossover. After the HRG vacuum subtraction, the large-volume result at $167$ MeV is $ 1.52(33)\times10^{-4}~{\rm GeV}^4$, compared with the quadratic contribution $ 1.00(16)\times10^{-4}~{\rm GeV}^4$. Thus, at the level of the central values, the net fourth-order contribution in this prescription amounts to roughly one half of the quadratic term. The fourth-order correction near the crossover therefore cannot be described as uniformly small. Because only two spatial volumes are
available, however, the large-volume point should not be interpreted as a controlled infinite-volume determination. At higher temperatures the situation is different. At $227$ MeV and above, the $O(B^4)$ contribution is small relative to the leading quadratic response, and the $O(B^2)$ approximation provides a good description of the magnetic pressure correction for $eB=0.2~{\rm GeV}^2$ at the precision of the present calculation. At lower temperatures the expansion parameter $eB/T^2$
grows rapidly, while the statistical uncertainties of $C_4$ also become large. Consequently, convergence of the Taylor series cannot be inferred from the small high-temperature values of $C_4$ alone. The pressure comparison therefore leads to the conclusion that the quadratic approximation is well supported at moderate and high temperatures for this field strength, but near the crossover the fourth-order response can be sizable and exhibits significant
sensitivity to the spatial volume. Additional volumes would be required to determine the thermodynamic-limit fourth-order pressure correction in this region.

\begin{table}[ht]
    \centering
    \caption{Magnetic contribution to the pressure at
    $eB=0.2~{\rm GeV}^2$ plotted on Fig.~\ref{fig:pressure}.
    All pressure values are given in units of
    $10^{-4}~{\rm GeV}^4$.  The first column gives the
    $O(B^2)$ contribution, the second includes the $O(B^4)$
    vacuum-unsubtracted term, and the third additionally subtracts the HRG estimate
    of the magnetic vacuum pressure.}
    \label{tab:pressure}
     \begin{threeparttable}
    \begin{tabular}{c c c c}
        \hline
        $T$ [MeV] &
        $10^4\Delta p_{O(B^2)}$ &
        $10^4\Delta p_{O(B^2)+O(B^4)}$ &
        $10^4\Delta p_{O(B^2)+O(B^4)-P_{\rm vac,HRG}}$ \\
        \hline
        134 & $-0.16(16)$ & $0.68(54)$  & $-0.12(54)$ \\
        154 & $ 0.24(16)$ & $0.98(39)$  & $ 0.18(39)$ \\
        167 & $ 0.96(20)$ & $1.47(33)$  & $ 0.67(33)$ \\
        167\tnote{1} & $ 1.00(16)$ & $ 2.32(33)$ & $1.52(33) $ \\
        173 & $ 1.36(20)$ & $1.86(28)$  & $ 1.06(28)$ \\
        227 & $ 4.12(32)$ & $4.17(33)$  & $ 3.37(33)$ \\
        373 & $ 7.72(52)$ & $7.73(52)$  & $ 6.93(52)$ \\
        611 & $11.08(56)$ & $11.08(57)$ & $10.28(57)$ \\
        \hline
    \end{tabular}
    \begin{tablenotes}
    \item[1]  Results using the large volume of $48^3\times 8$ for calculating $C_2^r$ and $C_4$.
    \end{tablenotes}
    \end{threeparttable}
\end{table}

\section{Computational aspects}

The principal computational advantage of our half-and-half Taylor approach is that all derivatives are evaluated at $B=0$. No new dynamical ensemble is required for each field value, in contrast with direct finite-$B$ scans. It is possible that at least through fourth order this method still has advantages, especially in combination with the minimal-noise link choice which reduces the stochastic variance of the determinant-derivative estimator.

The random-source counts in Table~\ref{tab:ensembles} should not be interpreted as the number required for $C_2^r$ alone. On the primary $32^3\times8$ ensembles around the crossover the extended calculation used 2400--4800 stochastic sources per configuration on the non-zero temperature ensembles because $C_4$ was being determined simultaneously. For the $48^3\times8$ finite-volume study at $T=167$ MeV, the statistics were increased much further, to between 9600 and 28800 stochastic sources per
configuration for the determination of the fourth-order coefficient. By contrast, the earlier $C_2$-only analysis already used 1200 stochastic sources at $167$, $173$, and $227$ MeV and obtained errors on $C_2^r$ comparable to those found after the later increases in stochastic statistics~\cite{LevkovaDeTar2014}. This indicates that at around 1200 sources the final uncertainty in $C_2^r$ was already close to a
gauge-noise floor. The much larger source count used in the $48^3\times8$ calculation should therefore be understood primarily in the context of resolving the considerably noisier fourth-order coefficient.
This distinction is important when comparing with later $B=0$ current-current methods: the $4.4\times10^5$ GPU-hour total is the cost of the enlarged $O(B^4)$ program, including finite-volume checks, not the cost of a dedicated susceptibility measurement.

\section{Retrospective context}

Lattice calculations using different methods confirmed that the thermal QCD medium is paramagnetic around and above the crossover and found weak diamagnetism at sufficiently low temperature \cite{Bali2014EOS,BaliEndrodiPiemonte2020}. These findings are consistent with our interpretation of Fig.~\ref{fig:C2r}. They also imply that our two low-temperature points, which are consistent with zero, lacked the precision to resolve the small negative susceptibility seen in later continuum studies.

A 2015 study by Bali and Endr{\H{o}}di examined the half-and-half construction for the bare zero-temperature susceptibility needed in a hadronic-vacuum-polarization application and found enhanced finite-volume effects associated with the discontinuous field profile \cite{BaliEndrodi2015HVP}. However, these effects cancel to a large extent in the thermal difference $\chi(T)-\chi(0)$. The latter is the quantity relevant to $C_2^r$ here. Therefore the finite-volume problem found for the unsubtracted zero-temperature susceptibility should not be transferred without qualification to the renormalized thermal susceptibility. The direct volume comparison in Fig.~\ref{fig:C2r} is consistent with that distinction.

In 2020, Bali, Endr{\H{o}}di, and Piemonte introduced a current-current-correlator method that also works entirely at $B=0$ and provides clean finite-volume control \cite{BaliEndrodiPiemonte2020}. For the thermal susceptibility alone, the conceptual difference is therefore not whether extra finite-$B$ ensembles are needed (neither approach needs them) but how the desired response is estimated. The half-and-half Taylor estimator builds the magnetic weighting directly into derivatives of the fermion matrix and uses a link representative selected to minimize stochastic noise; the current-current method reconstructs the relevant second moment from the electromagnetic correlator. No same-ensemble, equal-inversion-budget comparison has established which estimator has lower variance for $\chi(T)-\chi(0)$.

More recently, nonuniform magnetic fields have again been used as design tools for estimators. Brandt et al. determined the magnetic susceptibility from steady electric currents induced by a spatially varying magnetic field and found that the dominant valence contribution enables a comparatively inexpensive determination \cite{Brandt2024Steady}. This development is conceptually relevant to our minimal-noise construction: it reinforces the broader idea that the spatial form of the electromagnetic
source can be chosen not merely for formal convenience but also to improve the statistical efficiency of the estimators.

\section{Conclusions}

We have presented the fourth-order extension of the half-and-half Taylor calculation of the QCD pressure in an external magnetic field. The zero-net-flux field permits a well-defined differentiation at $B=0$, and the centered U(1) link phases were chosen explicitly to reduce stochastic noise. The updated $C_2^r$ results show a transition from values statistically consistent with zero at low temperature to a clearly positive thermal susceptibility above the crossover. A direct $32^3$-to-$48^3$ volume
comparison at 167~MeV shows no discernible finite-volume effect in this renormalized second-order coefficient.

The fourth-order coefficient is small and precisely determined at high temperature, while larger values occur in the crossover region. The updated finite-volume comparison at $T=167$ MeV gives $C_4(32^3\times8)=2.5(1.3)\times10^{-5}$, $C_4(48^3\times8)=6.4(1.4)\times10^{-5}$. The approximately $2\sigma$ difference provides evidence that finite-volume effects are more important for the unsubtracted $C_4$ than for $C_2^r$, for which no corresponding volume dependence is observed. Additional spatial volumes are required before a thermodynamic-limit value of the fourth-order coefficient can be determined near the crossover. For $eB=0.2~{\rm GeV}^2$, the $O(B^4)$ correction remains modest at moderate and high temperatures, supporting the quadratic approximation there. Near the crossover and at lower temperatures, however, the larger expansion parameter $eB/T^2$ and the observed volume sensitivity of $C_4$ imply that the quantitative fourth-order
pressure correction should be assigned an additional finite-volume uncertainty until a controlled volume study is performed.
The literature after 2014 confirms the qualitative paramagnetic high-temperature result and provides alternative estimators with cleaner systematic control. At the same time, the thermal subtraction substantially changes the relevance of the finite-volume criticism directed at the bare half-and-half susceptibility. The markedly different finite-volume behavior observed here for $C_2^r$ and $C_4$ also shows that assessments of the half-and-half method should distinguish between the
renormalized quadratic thermal susceptibility and higher-order response coefficients. A head-to-head variance-per-inversion comparison of the optimized half-and-half Taylor estimator with current-current and steady-current approaches for the thermal susceptibility would therefore be scientifically informative.

\section*{Acknowledgments}
This 2014 calculation used resources provided by the USQCD Collaboration, the University of Utah Center for High Performance Computing, and Indiana University. The underlying $O(B^2)$ work was supported in part by the U.S. National Science Foundation under Grant No.~PHY10-67881 and the U.S. Department of Energy under Grant No.~DE-FC02-12ER-41879 \cite{LevkovaDeTar2014}.

\printbibliography

\end{document}